\documentstyle[aps,preprint]{revtex}
\begin{document}
\draft

\preprint{KAIST-TH-97/3}

\title{Exactly solvable charged dilaton gravity theories
         in two dimensions}
\author{Youngjai Kiem$^{(a)}$, Chang-Yeong Lee$^{(b)}$ 
        and Dahl Park$^{(c)}$ \footnote{Corresponding author :
        Dahl Park \\
        E-mail : dpark@chep6.kaist.ac.kr \\
        Phone : 82-42-869-8132 \\
        Fax : 82-42-869-2510 } }
\address{$^{(a)}$ School of Physics,
                  Korea Institute for Advanced Study,
                  Seoul 130-012, Korea \\
         $^{(b)}$ Department of Physics,
                  Sejong University,
                  Seoul 143-747, Korea \\
         $^{(c)}$ Department of Physics,
                  KAIST,
                  Taejon 305-701, Korea \\
                  }
\maketitle

\begin{abstract}
We find exactly solvable dilaton gravity theories containing a
$U(1)$ gauge field in two dimensional space-time.  The classical 
{\em general} solutions for the gravity sector (the metric plus
the dilaton field) of the theories coupled to a massless complex 
scalar field are obtained in terms of the stress-energy tensor
and the $U(1)$ current of the scalar field.  We discuss issues
that arise when we attempt to use these models for the study 
of the gravitational back-reaction.
\end{abstract}


In black hole physics, the exactly solvable two dimensional 
model of Callan, Giddings, Harvey and Strominger (CGHS) played a 
significant role \cite{cghs}.  One of the main virtues of this 
model is its exact solvability and this aspect inspired many further
developments.  Using the exact classical solutions of the model, 
Schoutens, Verlinde and Verlinde concretely realized the black hole      
complementarity idea by mapping the dilaton gravity theory to
a critical string theory \cite{verlinde}.  
Balasubramanian and Verlinde performed a 
detailed analysis of the gravitationally dressed mass
shell dynamics using the CGHS model \cite{vijay}, based on the 
approach of Kraus and Wilczek \cite{per}.
The CGHS model captures the essential physics of the $s$-wave
Einstein gravity, while being a vastly more tractable model 
for analytical investigations.  

The main focus here is to find a class of two dimensional
dilaton gravity theories that include $U(1)$ gauge fields and,
at the same time, are simple enough to allow some analytic
treatment of their dynamics.  By studying these kinds of models
one hopes to further 
understand the black hole physics in extremal and
near-extremal regimes.  The dynamics of charged black holes
shows many of the intricate and interesting features.
In this regard, we mention that the recent advent of the D-brane
technology has provided us with a good motivation to better 
understand the (near) extremal black holes \cite{malda}.  
In D-brane approaches, 
by including a number of gauge fields, it is possible 
to consider the extremal limit where the Hawking temperature vanishes.
It is near this limit where most of the analysis were 
performed \cite{malda}.  In contrast, the Hawking 
temperature of the CGHS model is strictly a constant 
that is related to the cosmological constant.
In this note, we find one parameter class of exactly solvable  
two dimensional dilaton gravity models that contain a 
$U(1)$ gauge field and a complex charged scalar 
field \cite{dilaton}.  These models have the 
extremal limit with the vanishing Hawking temperature 
and, at the same time, retain the virtue of 
the CGHS model, i.e., the exact solvability of the 
gravity-dilaton sector.    

We consider the dilaton gravity theories given by the 
following action \cite{lowe}.
\begin{equation}
I = \int d^2 x \sqrt{-g} e^{-2 \phi} 
( R + \gamma g^{\alpha \beta} 
\partial_{\alpha} \phi \partial_{\beta} {\phi} + 
\mu e^{2 \lambda \phi}  -\frac{1}{4} e^{\epsilon \phi} F^2
\label{oaction}
\end{equation}
\[  - \frac{1}{2} e^{-2 ( \delta -1 ) \phi } 
g^{\alpha \beta} (\partial_{\alpha} -i e A_{\alpha} ) X
       (\partial_{\beta} +i e A_{\beta} ) X^*  )
\]
The fields in our consideration include a metric 
tensor $g_{\alpha \beta}$,
the dilaton field $\phi$, a $U(1)$ gauge field $A_{\alpha}$, and a 
massless complex scalar field $X$.  The scalar field $X$ has
the $U(1)$ charge $e$.  The signature choice for our metric
tensor is $(-+)$.  The two dimensional scalar 
curvature is denoted as $R$ and the $U(1)$ field strength is 
$F_{\alpha \beta} = \partial_{\alpha} A_{\beta} -
\partial_{\beta} A_{\alpha} $.  We use a convention $F^2 \equiv
g^{\alpha \beta} g^{\mu \nu} F_{\alpha  \mu} F_{\beta \nu}$.
A specific choice for real parameters $\gamma$, $\mu$, 
$\lambda$, $\delta$ and $\epsilon$ corresponds to a particular
dilaton gravity theory.  The CGHS model  \cite{cghs}, which has
been under intensive investigation partly due to their exact
solvability at the classical level, corresponds to the case when
$\gamma = 4$, $\mu > 0$, $\lambda = 0$, and $\delta = 0$.
The $s$-wave four dimensional Einstein gravity has the value
$\gamma = 2$, $\mu = 2$, $\lambda = 1$ and $\delta = 1$,
and when the electro-magnetic fields are included, $\epsilon = 0$
\cite{source}.
Although its importance is significant, the four dimensional 
Einstein gravity has not been solved exactly even at the $s$-wave
level, which provides some motivation for devising a simple
and exactly solvable model theory of gravity that mimics its
physical properties.

The algebraic properties of the curvatures
are simple in two dimensions.  For example, we have   
$R_{\alpha \beta} - g_{\alpha \beta} R/2 = 0$ and
$g_{\alpha \beta} F^2 = 2 g^{\mu \nu} F_{\alpha \mu} F_{\beta \nu}$
as algebraic identities.  These identities simplify the actual 
calculations.   Additionally, the behavior of the gravity sector, 
which includes the graviton and
the dilaton, and the $U(1)$ gauge field sector is particularly simple
in two dimensional space-time.  The metric tensor and the dilaton field
have the total of four independent components.  Recalling that we have
two reparameterization invariance $x^{\alpha} \rightarrow x^{\alpha}
+ \epsilon^{\alpha} ( x^{\alpha}) $, the physical propagating degrees
of freedom for the gravity sector is absent.   Likewise, the gauge field
has two components and the existence of the $U(1)$ gauge symmetry
suggests that there are also no physical propagating degrees of freedom
for this sector. 

The equations of motion from the action, Eq. (\ref{oaction}), are 
given by
\begin{equation}
- D_{\alpha} D_{\beta} \Omega + g_{\alpha \beta} D \cdot D \Omega
- \frac{\gamma}{8} ( g_{\alpha \beta}  
\frac{(D \Omega )^2}{\Omega} - 2 \frac{D_{\alpha } \Omega 
D_{\beta} \Omega }{\Omega} ) 
\label{eom1}
\end{equation}
\[ - \frac{\mu}{2} g_{\alpha \beta} 
\Omega^{1- \lambda} - \frac{1}{8} g_{\alpha \beta} F^2
\Omega^{1 - \epsilon / 2 } = T^X_{\alpha \beta} , \]
\begin{equation}
R + \frac{\gamma}{4} ( \frac{(D \Omega )^2 }{\Omega^2} 
  - 2 \frac{D \cdot D \Omega}{\Omega} ) 
+ (1- \lambda ) \mu \Omega^{- \lambda} -
\frac{1}{4} (1 - \frac{\epsilon}{2} ) \Omega^{-\epsilon /2 } F^2 =
T^{X}_{\Omega}  ,
\label{eom2}
\end{equation}
\begin{equation}
  g^{\nu \beta } ( (1 - \frac{\epsilon}{2} ) 
\frac{D_{\nu} \Omega}{\Omega} F_{\alpha \beta} 
+ D_{\nu} F_{\alpha \beta} )
= j^{X}_{\alpha} ,
\label{eom3}
\end{equation}
and
\begin{equation}
 (D_{\alpha} + ie A_{\alpha} ) ( g^{\alpha \beta} \Omega^{\delta} 
( D_{\beta} + ie A_{\beta} ) X^* ) = 0 
\label{eom4}
\end{equation}
along with its complex conjugate.  We introduce 
$\Omega = e^{ - 2 \phi }$ and $D_{\alpha}$ denotes a covariant
derivative, along with $D \cdot D = g^{\alpha \beta} D_{\alpha}
D_{\beta} $, the two dimensional Laplacian.  The stress-energy
tensor of the scalar field $X$ is given by
\begin{equation}
T^X_{\alpha \beta}
=  \frac{\Omega^{\delta}}{2} (
  (\partial_{\alpha} - ie A_{\alpha} ) X 
      (\partial_{\beta} + ie A_{\beta} ) X^* 
- \frac{1}{2} g_{\alpha \beta} g^{\mu \nu}
( \partial_{\mu} - ieA_{\mu} ) X (\partial_{\nu} + ieA_{\nu} ) X^*),
\label{tab}
\end{equation} 
and the dilaton charge and the $U(1)$ current of the scalar
field are
\begin{equation}
T^{X}_{\Omega} = \frac{\delta}{2} \Omega^{\delta -1} 
g^{\mu \nu} (\partial_{\mu} - ieA_{\mu} ) X
                    (\partial_{\nu} +ie A_{\nu} ) X^*
\label{tomega}
\end{equation}
\begin{equation}
j^{X}_{\alpha} = \frac{1}{2} \Omega^{\delta -1 + \epsilon /2 }
( ie ( X \partial_{\alpha} X^* - X^* \partial_{\alpha} X )
 - 2 e^2 A_{\alpha} X X^* ) .
\label{u1current}
\end{equation}
We note that, due to the absence of the mass term for the scalar field,
the stress-energy tensor $T^X_{\alpha \beta}$ is traceless, i.e.,
$g^{\alpha \beta} T^X_{\alpha \beta} = 0$.

The key simplification of the equations of motion occurs when
we rescale the metric via
\begin{equation}
g_{\alpha \beta } \rightarrow \Omega^{- \gamma /4 } g_{\alpha \beta} .
\label{rescale}
\end{equation}
This transformation cancels the kinetic term for the dilaton field
in the action, Eq. (\ref{oaction}), up to total derivative terms.  
The remaining changes in Eq. (\ref{oaction}) can be summarized
formally by 
\[ \epsilon \rightarrow \epsilon - \frac{\gamma}{2} 
\ \ \ , \ \ \  \lambda \rightarrow \lambda + \frac{\gamma}{4} .\]
Thus, the equations of motion are simplified to be
\begin{equation}
- D_{\alpha} D_{\beta} \Omega + g_{\alpha \beta} D \cdot D \Omega
 - \frac{\mu}{2} g_{\alpha \beta} 
\Omega^{1- \lambda - \gamma / 4} - \frac{1}{8} g_{\alpha \beta} F^2
\Omega^{1 - \epsilon / 2 + \gamma / 4 } = T^X_{\alpha \beta} , 
\label{new1}
\end{equation}
\begin{equation}
R
+ (1- \lambda - \frac{\gamma}{4} ) \mu \Omega^{- \lambda -\gamma /4 } 
- \frac{1}{4} (1 - \frac{\epsilon}{2} + \frac{\gamma}{4}  ) 
\Omega^{-\epsilon /2 + \gamma /4} F^2 =
T^{X}_{\Omega}  ,
\label{new2}
\end{equation}
\begin{equation}
  g^{\nu \beta } ( (1 - \frac{\epsilon}{2} + \frac{\gamma}{4} ) 
\frac{D_{\nu} \Omega}{\Omega}  F_{\alpha \beta} 
+ D_{\nu} F_{\alpha \beta} )
= j^{X}_{\alpha} .
\label{new3}
\end{equation}
The stress-energy tensor $T^X_{\alpha \beta}$  and $T^X_{\Omega}$
do not change, while we have a modified expression for
$j^{X}_{\alpha}$ as;
\begin{equation}
j^{X}_{\alpha} = \frac{1}{2} \Omega^{\delta -1 + \epsilon /2 - \gamma / 4}
( ie ( X \partial_{\alpha} X^* - X^* \partial_{\alpha} X )
 - 2 e^2 A_{\alpha} X X^* ) .
\label{newu1}
\end{equation}
The equations of motion for the scalar field $X$, Eq. (\ref{eom4})
and its complex conjugate,
do not change under the rescaling of the metric.

With equations of motion written in the above fashion, we observe that
imposing the conditions for the parameters
\begin{equation}
  1- \lambda - \frac{\gamma }{4} = 0 \ \ \ ,
   \ \ \ 1 - \frac{\epsilon}{2} + \frac{\gamma}{4} = 0
   \ \ \ , \ \ \ \delta = 0
\label{pcon}
\end{equation}
renders exactly solvable theories.  We note that in the absence of the
$U(1)$ gauge field, the CGHS model satisfies the condition
$ 1- \lambda - \gamma / 4 = 0$ and $\delta = 0$.  For the choice
of parameter values for the CGHS model, the above conditions
determine $\epsilon = 4$.  The consequences of the conditions 
are manifold.  In Eq.(\ref{new1}), the non-linear terms for
$\Omega$ field are absent on the left hand side and 
$T^X_{\alpha \beta}$ does not depend on the $\Omega$ field.
Additionally, $j^X_{\alpha}$ also does not depend on the $\Omega$
field, and $T^{X}_{\Omega}$ vanishes identically.  We have
\begin{equation}
-D_{\alpha} D_{\beta} \Omega + g_{\alpha \beta} D \cdot D
\Omega - ( \frac{\mu}{2} + \frac{1}{8} F^2 ) g_{\alpha \beta}
= T^X_{\alpha \beta} 
\label{moe1}
\end{equation}
\begin{equation}
R = 0
\label{moe2}
\end{equation}
\begin{equation}
g^{\mu \nu} D_{\mu} F_{\alpha \nu} = j^X_{\alpha}
\label{moe3} .
\end{equation}
Eq. (\ref{moe2}) implies that the metric $g_{\alpha \beta}$
is flat.  We can thus choose to use the flat conformal 
coordinates $ds^2 = - dx^+ dx^-$ for the description of the
space-time geometry.  Under this gauge choice for the 
space-time coordinates, we can rewrite Eq.(\ref{moe1}) as
\begin{equation}
\partial^2_+ \Omega = - T^X_{++} \ \ \ , \ \ \
\partial^2_- \Omega = - T^X_{--} \ \ \ , \ \ \
\partial_+ \partial_- \Omega = - \frac{\mu}{4} + \frac{f^2}{2} ,
\label{con1}
\end{equation}
where we introduce $F_{+-} = f = - F_{-+} $ and we use the fact
that the trace of $T^X_{\alpha \beta}$ vanishes.
Eq. (\ref{moe3}) becomes
\begin{equation}
2 \partial_+ f = - j^X_+ = -ie (X \partial_+ X^* - X^* 
                 \partial_+ X ) + 2 e^2 A_+ X X^*
\label{con2}
\end{equation}
\[ 2 \partial_- f = j^X_- = ie (X \partial_- X^* - X^* 
                 \partial_- X ) - 2 e^2 A_- X X^* . \]
The scalar field $X$ satisfies
\begin{equation}
(\partial_+ + ie A_+ ) ( (\partial_- + ie A_- ) X^* ) +
(\partial_- + ie A_- ) ( (\partial_+ + ie A_+ ) X^* ) = 0 ,
\label{con3}
\end{equation}
from Eq. (\ref{eom4}), and its complex conjugate.
The crucial point is that Eqs. (\ref{con2}) and (\ref{con3})
are identical to the ones in the flat space-time scalar 
electrodynamics, and they can be solved to
determine $A_{\pm}$ and $X$ with proper 
initial and boundary conditions, up to gauge transformations.  
This will in turn determine $T^X_{\pm \pm}$ via
\begin{equation}
T^X_{\pm \pm} = \frac{1}{2} (\partial_{\pm} - ie A_{\pm} ) X
   (\partial_{\pm} + ie A_{\pm} ) X^* .
\end{equation}
We can consistently integrate both Eq. (\ref{con1}) and
Eq. (\ref{con2}), once we determine $j^X_{\pm}$ and
$T^X_{\pm \pm}$.  Via the equation of motion for the $X$ field, 
Eq. (\ref{con3}), we see that the integrability condition for
Eqs. (\ref{con2}),
\begin{equation}
\partial_+ j^X_- + \partial_- j^X_+  = 0, 
\label{int1}
\end{equation}
which represents the $U(1)$ current conservation (all
other fields in our consideration being neutral), holds.
Similarly, upon using Eqs. (\ref{con2}) and (\ref{con3}),
we verify that the integrability conditions
\begin{equation}
 \partial_- T^X_{++} = - f \partial_+ f  \ \ \ ,
 \ \ \  \partial_+ T^X_{--} = - f \partial_- f 
\label{int2}
\end{equation}
for Eqs. (\ref{con1}) are satisfied.  Eqs. (\ref{int2}) represent 
the stress-energy conservation for the scalar field $X$ and
the $U(1)$ gauge field.  We note that in the presence of
non-vanishing charge currents, $T^X_{\pm \pm}$ are not
chirally conserved.  

The general solution for Eq. (\ref{con2})
is solved to be
\begin{equation}
f (x^+ , x^- ) =  f( x_0^+ , x_0^- ) -
    \frac{1}{2} \int_{x^+_0}^{x^+} j^X_+ 
    ( x^{+ \prime} , x^-_0 ) d x^{+ \prime} 
    +  \frac{1}{2} \int_{x^-_0}^{x^-} j^X_- 
    ( x^+ , x^{- \prime}  ) d x^{- \prime}     
\label{fsol}
\end{equation}
where $x_0^{\pm}$ are arbitrary constants.
This solution determines the $U(1)$ field strength $f$
in terms of the incoming ($j^X_+$) and the outgoing
($j^X_-$) charge currents.  The general solution
for Eq. (\ref{con1}) is
\begin{equation}
\Omega (x^+ , x^- )
= \Omega (x^+_1 , x^-_1 ) 
- \frac{\mu}{4} (x^+ - x^+_1 ) ( x^- - x^-_1 )
\label{osol}
\end{equation}
\[ + \frac{1}{2} \int_{x^+_2}^{x^+} \int_{x^-_2}^{x^-}
f^2 ( x^{+ \prime} , x^{- \prime} ) dx^{- \prime} dx^{+ \prime}
 - \frac{1}{2} \int_{x^+_2}^{x^+_1} \int_{x^-_2}^{x^-_1}
f^2 ( x^{+ \prime} , x^{- \prime} ) dx^{- \prime} dx^{+ \prime} \]
\[ - \int_{x^+_1}^{x^+} \int_{x^+_3}^{x^{+ \prime}}
   T^X_{++} ( x^{+ \prime \prime} , x^-_2 ) 
    dx^{+ \prime \prime} dx^{+ \prime}
   - \int_{x^-_1}^{x^-} \int_{x^-_3}^{x^{- \prime}}
   T^X_{--} ( x^+_2 , x^{- \prime \prime}  ) 
    dx^{- \prime \prime} dx^{- \prime}  \]
where $x_1^{\pm}$, $x_2^{\pm}$ and $x_3^{\pm}$ are
arbitrary constants.  The dilaton field is determined
in terms of the scalar field stress-energy tensor
$T^X_{\pm \pm}$ and the $U(1)$ field strength, which,
in turn, is given in Eq. (\ref{fsol}) via the 
charge currents of the scalar field.  These are the main
results of this note.  Now that we solved the dilaton
and the metric in terms of matter stress-energy tensor
and the $U(1)$ current, we reduced our problem to that
of the two dimensional scalar electrodynamics in flat
space-time.

Static solutions of the theories in our consideration 
have been discussed in the literature \cite{mass}.  
The particularly
simple case is when we have the gravity-dilaton sector
of the CGHS model.  Thus, in our further consideration
we set $\gamma = 4$.  The static solutions in this case
are given by
\begin{equation}
\Omega = e^{- 2 \phi} = \Omega_0 
- (\mu / 4 - f_0^2 /2 ) x^+ x^-
\label{dil}
\end{equation}
for the dilaton field and
\[ ds^2 = - dx^+ dx^- / \Omega \]
for the metric.  Here $f_0$ is the charge of the $U(1)$
gauge field.  This solution exhibits the general feature
of the higher dimensional stringy charged black holes.
If the magnitude of the charge $f_0$ is less than the
critical value $ \sqrt{\mu /2}$, we have a black hole
geometry with one horizon, which is at $x^- = 0$.
Hidden inside this horizon, we have the curvature singularity
at $x^+ x^- =  2 \Omega_0 / ( \mu /2 - f_0^2 )$.
Unlike the Reissner-Nordstrom black holes, the stringy
charged black hole has a curvature singularity at the
would-be inner horizon, and this is the property that our
solutions share \cite{horowitz}.  
If $f_0^2 = \mu /2 $, we have
space-time geometries that have flat incoming region and
flat outgoing region.  This behavior is also similar to 
the higher dimensional stringy charged black holes where
extremal solutions have a long throat that connects two
flat space-time regions \cite{horowitz}.  
If $f_0^2 > \mu /2 $, we have
naked singularities.  This consideration suggests that
we can regard $ f_0^2 = \mu /2 $ case as extremal.
Indeed, by computing the period of the Euclideanized
time coordinate, we determine the Hawking temperature
to be 
\begin{equation} 
T_H = \frac{1}{2 \pi} \sqrt{  \frac{\mu}{4} 
 - \frac{f_0^2}{2} } 
\end{equation}
and the mass of the black hole is given by
\begin{equation}
M = \Omega_0 \sqrt{ \frac{\mu}{4} - \frac{f_0^2}{2} } .
\end{equation}
Under the suggested condition of the extremality, 
we find the Hawking temperature vanishes.  This is a 
generic behavior of charged extremal black holes.   

The usual way to compute the mass is to integrate
the perturbed metric over a space-like hypersurface.
However, we can demonstrate the validity
of the mass expression given 
above by considering the injection of  
charge-neutral stress-energy flux.  For this purpose,
we rewrite the dilaton expression as
\begin{equation}
\Omega = ( \Omega_0 + \int^{x^+} dx^{+ \prime} x^{+ \prime} 
T^X_{++} )  - k^2 x^+ ( x^- + \frac{1}{k^2}
 \int^{x^+}  T^X_{++} dx^{+ \prime} )
\label{ttt}
\end{equation}
where we set $T^X_{--} = 0$, use an integration 
by parts and introduce $k = \sqrt{ \mu /4 - f_0^2 /2 } $.  
Since we consider the neutral scalar field,
the black hole charge $f_0$ does not change.   
The asymptotically flat coordinates $v$ and $u$
are related to the Kruskal-like coordinates $x^{\pm}$
via 
\[ v = \frac{1}{k} \ln x^+ \ \ \ , \ \ \ 
   u = - \frac{1}{k} \ln ( -x^- )   \]
outside the black hole, and we see that 
\[ \int^{x^+} dx^{+ \prime} x^{ + \prime} 
  T^{X}_{++} = \frac{1}{k} \int^{v} dv T^X_{vv} .\] 
Since the mass is in fact $\Omega_0 = M / k$, we find that
the first part of Eq.(\ref{ttt}) denotes the dynamical
increase of black hole mass by the injection of the
matter stress-energy.  The second part of Eq.(\ref{ttt})
shows the shift interaction considered 
by 't Hooft \cite{thooft}.

Given our results on the classical back-reaction on
the space-time caused by the propagating charged scalar fields
as shown in Eqs. (\ref{fsol}) and (\ref{osol}), 
the natural next step is to use the gravity theories
in this note for the investigation of the quantum 
gravitational back-reaction.  The crucial issue then
is the introduction of
a boundary.  The same situation happens
in the quantization of the CGHS model by Schoutens,
Verlinde and Verlinde \cite{verlinde}.  To mimic the dynamics
of the four dimensional $s$-wave Einstein gravity, they 
introduce a reflecting boundary that plays the role
of $r = 0$ in the $s$-wave Einstein gravity.  We note that
this also 
allows one to avoid the strong coupling region to be 
asymptotic in- or out-regions.  Once this boundary
is introduced, the properties of the back-reaction in 
the CGHS model become qualitatively similar to those of 
the $s$-wave Einstein gravity \cite{kvv}.  Ultimately, one 
hopes to relate the behavior of the models considered
here to the $s$-wave reduction of the five dimensional
(or four dimensional) supergravity theory \cite{comment}.  
We note that
similar to the $s$-wave reduction of the Einstein gravity,
we wrote down the two dimensional $s$-wave action of the 
type IIB supergravity on $T^5$ in
Ref \cite{klp}. 
In other words, 
our models may share the same qualitative 
properties as those
of the $s$-wave supergravity theories when an interacting
boundary is introduced (including the gauge interactions).  
In the context of
the CGHS model, the dynamical moving mirror of 
Ref. \cite{mirror} provides one such example.  In view
of the recent developments involving D-brane technology,
the following possibility comes out.  The important 
dynamics of the D-brane approach takes place at the level
of $s$-wave.  It is a tempting idea, then, to have a description
in terms of an effectively two dimensional field theory with
non-trivial boundary interactions, rather similar to
the Callan-Rubakov effect.  In this picture, it may 
be possible that we can
regard the whole D-brane configurations at weak
coupling as a boundary point where we impose some form of
the reflecting boundary condition.  As the loop correction
makes the D-branes massive, we expect quite non-trivial
dynamics of the boundary point itself.  Although our models
are {\em different} from the low energy effective theory of 
the D-brane systems, we expect that they may provide 
an insight along this direction in a simpler setting.  
We plan to discuss the (semi-classical) quantization of our 
model theories in the future.

\acknowledgements{Y. K. and C.Y. L. were supported in part
by the Basic Science Research Institute Program, Ministry of
Education, BSRI-96-2442.  D. P. would like to thank Prof.
Jae Kwan Kim for helpful guidance and discussions.
We also thank W.T. Kim and H. Verlinde for useful 
discussions.}

\appendix


\begin{thebibliography}{99}
\bibitem{comment} It is easy to generalize our work to include
      an arbitrary number of $U(1)$ gauge fields and scalar fields.
\end{thebibliography}
\end{document}